\documentclass[12pt]{article}

\usepackage{amsfonts,amsmath,amsxtra}

\newcommand{\beq}[1]{\begin{equation}\label{#1}}
\newcommand\eeq {\end{equation}}
\newcommand\bqa {\begin{eqnarray}}
\newcommand\eqa {\end{eqnarray}}

\newcommand{\eq}[1]{eq.(\ref{#1})}
\newcommand{\bear}{\begin{array}}
\newcommand{\enar}{\end{array}}

\newcommand{\A}{\mathbb{A}}

\newcommand{\cP}{\mathbb{P}}

\begin{document}
\def\t{\theta}
\def\T{\Theta}
\def\w{\omega}
\def\ov{\overline}
\def\a{\alpha}
\def\b{\beta}
\def\g{\gamma}
\def\s{\sigma}
\def\l{\lambda}
\def\wt{\widetilde}

\hfill{ITEP--TH--49/06}
\vspace{10mm}

\centerline{\bf \Large On the relation between Unruh and
Sokolov--Ternov effects}
\vspace{7mm}
\centerline{{\bf Emil T.Akhmedov}\footnote{akhmedov@itep.ru}}
\centerline{Moscow, B.Cheremushkinskaya, 25, ITEP, Russia 117218}
\vspace{3mm}
\centerline{and}
\vspace{3mm}
\centerline{{\bf Douglas Singleton}\footnote{dougs@csufresno.edu}}
\centerline{Physics Department, CSU Fresno, Fresno, CA 93740-8031}
\vspace{10mm}
\begin{abstract}
We show that the Sokolov--Ternov effect --- the depolarization of
particles in storage rings coming from synchrotron radiation due
to spin flip transitions --- is physically equivalent to the Unruh
effect for circular acceleration if one uses a spin 1/2 particle
as the Unruh--DeWitt detector. It is shown that for the electron,
with gyromagnetic number $g \approx 2$, the
exponential contribution to the polarization, which usually
characterizes the Unruh effect, is
``hidden" in the standard Sokolov-Ternov effect making
it hard to observe. Thus, our conclusions are different in
detail from previous work.
\end{abstract}
\vspace{10mm}

\section{Introduction}

One of the goals of this note is to show that the exotic effect of particle
observation by a detector undergoing a non--inertial motion can be seen as
an ordinary quantum field theory (QFT) phenomenon. This is done in detail
for
the Unruh effect for circular motion but by implication should extend
to other situations, e.g. the Hawking radiation from a black hole.
We do not derive any new formula, but rather present well known ones in a
new light
and establish relations between seemingly unrelated phenomena.

In the Unruh effect a classical detector moving with acceleration
in the background of a QFT can excite its internal degrees of
freedom even if the QFT at the beginning is in its ground state.
The standard Unruh effect is the situation where a detector moves
eternally with constant acceleration (in direction and magnitude)
and excites its internal degrees of freedom \cite{Unruh}. In this
work by the Unruh effect we mean a more general phenomenon: a
detector in {\it any} non--inertial, homogeneous motion in {\it
flat, empty Minkowski space} which excites its internal degrees of
freedom. Note that the excitation which is of interest to us is
not due to the fact that one part of the detector moves with
respect to another one (like the pointer/arrow of an ammeter which
moves with respect to its box if it is shaken), but rather is due
to the existence of radiation in the detector's non--inertial
reference frame. If the detector undergoes {\it homogeneous}
motion, then all its parts are in static equilibrium with respect
to each other and shaking/excitation due to internal forces does
not occur. Homogeneous motions happen under (combined) actions of
two constant forces --- constant in direction and magnitude
(causing linear acceleration) and constant in magnitude (causing
orbiting acceleration). The excitation of the detector  due to its
shaking it is not a universal phenomenon, but depends on its
internal structure. It is for this reason that we consider
homogeneous motion, since then every thing just depends on the
features of the radiation detected by the detector in its
non--inertial reference frame. As we will see the characteristic
features of the radiation in question depend on the type of
homogeneous motion performed by the detector. Whether the
radiation is thermal or not is {\it not} important to us in this
note. We confirm previous observations that the radiation in the
orbiting reference frame is non--thermal which contradicts some
statements of \cite{Unruh}. What is important for us in this note
is simply the existence of the radiation, which shows that
detectors in linear accelerating motion and in orbiting motion get
excited for the same physical reason --- due to the existence of
the radiation detected by the detector in its non--inertial
reference frame.

Qualitatively this effect is very similar to the Hawking radiation
of black holes \cite{Hawking} (see the discussion in the
concluding section). Normally these effects are too small to be
detected experimentally, with the one possible exception being the
radiation detected by a detector in uniform circular motion. Thus
we examine to what extent it might be possible to detect the Unruh
type characteristic exponential factor in the excitation
probability rate for circular motion. It is shown that for
electrons with gyromagnetic number $g \approx 2$ that the
characteristic exponential behavior of the excitation probability
rate is ``hidden" in the standard Sokolov--Ternov effect and
is thus hard to observe. However, we show that particles with
general $g$ do experience substantial exponential factor
contribution. Thus, another goal of this paper is to investigate
the question of the experimental observability of the
characteristic Unruh type exponential factor in the excitation
probability rate.

In this paper in all cases we consider Minkowski spacetime, and
we take $\hbar =1$ and $c=1$. We study the reaction of detectors in
motion with respect to vacuum fluctuations of a background scalar quantum
fields. For simplicity we consider a two--level detector. The specific
example we have in mind is a spin $1/2$ particle in a magnetic field.
Our considerations are easy to generalize to detectors
with an arbitrary number of states and also background quantum fields other
than scalar.
The energy of the ground state of the detector, $|0\rangle$, is
$\mathcal{E}_d$, while the energy of
its excited state, $|1\rangle$, is $\mathcal{E}_{u}$. The
characteristic monopole interaction Hamiltonian for such a detector in
background scalar fields $\phi_a$, $a=1,\dots, N$ is
\bqa
\hat{H}_{int}(t) = q\,\sqrt{\dot{x}_\mu^2(t)} \, \hat{\mu}_a[x(t)]\,
\hat{\phi}_a[x(t)] =
q \, \sqrt{\dot{x}_\mu^2(t)} \, e^{-{\rm i}\, \hat{P}_\mu^{(D)} \,x^\mu(t)}
\, \hat{\mu}_a(0)\,
e^{{\rm i}\,
\hat{P}_\mu^{(D)} \,x^\mu (t)} \,\hat{\phi}_a[x(t)], \label{bla}
\eqa
where $\hat{\mu}_a$ is the detector's monopole moment, which describes its
interaction
with the external scalar fields $\phi_a$; $\hat{P}_\mu^{(D)}
= (\hat{H}_D, \, \hat{\vec{P}}_D)$ is the detectors four--momentum
governing the dynamics of the free detector (in the detector's
rest frame $\hat{H}_D |0\rangle = \mathcal{E}_{d} |0\rangle$ and $\hat{H}_D
|1\rangle = \mathcal{E}_{u} |1\rangle$); $x(t)$ is the detector's
trajectory; $q$ is the coupling constant; $\dot{x}_\mu^2(t) =
g_{\mu\nu}(x)\,
(dx^{\mu}/dt) \, (dx^{\nu}/dt)$, where $g_{\mu\nu}$ is the metric of the
space--time. It is easy to see that the
Hamiltonian in question is invariant under general covariance and
re-parameterization
invariance (change of $t$). Below we consider situations where
$\sqrt{\dot{x}_\mu^2(t)}$
is equal either 1 or $\gamma^{-1}$ depending on whether we consider $t$ as
proper or laboratory
time; $\gamma$ is the relativistic $\gamma$--factor. Furthermore, if we
consider the detector's rest frame then
$\hat{P}_\mu^{(D)} \,x^\mu(t) = \hat{H}_D \, t$, where $t$ is the proper
time. In all analysis below
we will indicate whether we are using proper or laboratory time.

To understand the origin of this Hamiltonian recall the one
describing the standard non--relativistic interaction of the spin
$\vec{s}$ with the magnetic field $\vec{H}$: $\hat{H}_{int}
\propto \vec{s}[t]\cdot \vec{H}[x(t)]$. In \eq{bla} we just
consider the simplified version of such an interaction between the
relativistically moving detector, (this is the reason for the
factor $\sqrt{\dot{x}_\mu^2}$)  and scalar fields $\phi_a$ instead
of a vector field $\vec{H}$ (this is the reason for the factor
$\hat{\mu}_a$ instead of spin $\vec{s}$).

In the main case of interest to us both the detector and the background QFT
are originally in their ground states and the detector moves along a
specified
trajectory $x(t)$. We want to find the probability
for the detector to get excited at the cost of work performed by
the force pulling the detector along the trajectory.
As the result of such a process the background QFT
will become excited as well.

To leading order in $q$ the amplitude for such a process is:
\bqa
\A = {\rm i} \, \int dt \, \left\langle 1 \left| \otimes \left\langle \psi
\left|
\hat{H}_{int}^{\phantom{\frac12}}(t) \right| {\rm Vac}\right\rangle \otimes
\right|
0\right\rangle,
\eqa
where $|\psi\rangle$ is the final state of
the QFT. Hence, to leading order in
perturbation theory the probability in question in the co-moving
reference frame is:
\bqa
P \propto q^2 \,
\left\langle 0 \left| \mu^{\phantom{\frac12}}_a(0)\right|1 \right\rangle
\left\langle 1\left| \mu^{\phantom{\frac12}}_b(0)\right|0 \right\rangle \,
\int \int dt \, dt'\, e^{- {\rm i}\,
\Delta \mathcal{E}\, \left(t - t'\right)} \, \left\langle 0\left|
\phi^{\phantom{\frac12}}_a[x(t)] \right|\psi \right\rangle \,
\left\langle \psi\left| \phi^{\phantom{\frac12}}_b [x(t')]
\right|0 \right\rangle, \label{grule1} \eqa where $\Delta
\mathcal{E} = \mathcal{E}_u - \mathcal{E}_d$. A non--vanishing
contribution for this expression appears when $|\psi\rangle$
contains one quantum of at least one of the fields $\phi_a$.

We would like to consider the total probability rate per unit
time. To obtain it we have to sum \eq{grule1} over all final
states $\psi$ of the QFT using the decomposition: \bqa \sum_{\psi}
|\psi \rangle \, \langle \psi | = 1.\label{psisum} \eqa Because
the integral over $t$ in \eq{grule1} has not yet been taken,
energy conservation is not imposed. Hence, in \eq{psisum} we are
summing over {\it all} states of the QFT rather than over those at
a given energy. Now we change the integration variables in
\eq{grule1} to $\tau = t - t'$ and $\tau' = t + t'$ and dropping
the integral over $\tau'$ we obtain the probability decay rate in
question:
\bqa w(t) \propto q^2 \, \left\langle 0\left|
\mu_a^{\phantom{\frac12}}(0)\right|1\right\rangle \, \left\langle
1\left| \mu_b^{\phantom{\frac12}} (0)\right|0\right\rangle \,
\int_{-\infty}^{+\infty} d\tau \, e^{- {\rm i}\, \Delta
\mathcal{E}\, \tau}\, G_{ab}\left[x(t -
\tau/2)^{\phantom{\frac12}},\,\,x(t + \tau/2)\right],
\label{grule2}
\eqa
where $G_{ab}\left[x(t - \tau/2),\,x(t +
\tau/2)\right] = \left\langle \phi_a[x(t -
\tau/2)]^{\phantom{\frac12}} \phi_b[x(t + \tau/2)]\right\rangle$
is the Wightman function of the QFT. This derivation of
\eq{grule2} from \eq{grule1} is correct only for homogeneous
motions (for a more detailed study see \cite{Schlicht}).

In more general situations the state of the background QFT is
characterized by the density matrix $\hat{\rho}$. For the thermal state
\bqa
\rho_{MK} = \delta_{MK}\,
\frac{e^{ - \frac{\mathcal{E}_M}{T}}}{Z_{tot}}, \quad {\rm and} \quad
Z_{tot} = \sum_M e^{- \frac{\mathcal{E}_M}{T}},
\eqa
where $M$ and $K$ are multi--indices
labeling the energy levels of the QFT. In the zero temperature
limit, $T\to 0$, only the vacuum state survives and we go back to the
situation described in the previous paragraph.

Additionally we want to find the probability rate for the
detector to change from the ground state to the excited one and vise versa.
A derivation similar to the one above with finite temperature leads to the
following golden rule formula for this probability rate:
\bqa
w_{\mp}(t) \propto  q^2 \,\left\langle
0\left| \mu_a^{\phantom{\frac12}}(0)\right|1\right\rangle \,
\left\langle 1\left| \mu_b^{\phantom{\frac12}}(0)\right|0\right\rangle\,
\int_{-\infty}^{+\infty} d\tau \, e^{\mp
{\rm i}\, \Delta \mathcal{E}\, \tau}\,
G_{ab}^{(T)}\left[x(t-\tau/2)^{\phantom{\frac12}}, \,x(t+\tau/2)\right],
\label{grule3}
\eqa
where now $G_{ab}^{(T)}\left[x(t-\tau/2)^{\phantom{\frac12}},
\,x(t+\tau/2)\right]
= {\rm Tr} \, \left\{\hat{\rho}\, \phi_a[x(t-\tau/2)]^{\phantom{\frac12}}
\phi_b[x(t+\tau/2)]\right\}$ is the thermal Wightman function and the trace
is taken over all states of the QFT. In this equation the ``$-$'' sign
corresponds to the rate for the detector to get excited from
its ground state, while the ``$+$'' sign corresponds to the
rate of the excited detector to decay to the ground state.

There are several well understood situations in which
\eq{grule3} gives non--zero transition rates. We will
study most of these in detail in the main body of the text. The first
example we will examine is a static detector in the heat bath. In this
case \eq{grule3} yields the standard Planckian behavior
for $w_{\pm}$, i.e. the detector sees particles with a Planckian spectrum.
Furthermore, if one considers \eq{grule3} at
$T=0$ one finds $w_{-} = 0$, but $w_{+}\neq 0$ since there is zero
probability
for the detector to get excited in this case, but if
the static detector is originally in its excited state there is
a non--zero probability for it to {\it spontaneously} decay to the ground
state.

A more exotic situation is the Unruh effect \cite{Unruh}. To
describe this effect in the standard way one considers the
detector moving eternally with constant (in direction and
magnitude) acceleration. Additionally one takes the internal
degrees of freedom of the detector as decoupled from the external
ones as can be seen from the derivation of \eq{grule2}. Let us
explain this point. In the standard description of the Unruh
effect there are three forces which in general are of a different
nature. The first force pulls the detector along its trajectory.
The second force is associated with the energy splitting between
$\mathcal{E}_{d}$ and $\mathcal{E}_{u}$. The third force
corresponds to the fields which are radiated when the detector
changes its internal state. This situation is very hard to realize
experimentally. For a charged particle in an accelerator storage
ring all of these three forces are electromagnetic in nature. It
is this fact which allows one to see the Sokolov--Ternov effect as
the Unruh effect but ``spoiled'' by several impurities.

In section 3 we show at $T=0$ and with $x(t)$
describing hyperbolic motion (i.e. constant acceleration and
$t$ being the proper time of the detector), that \eq{grule3} yields a
detector
which sees particles of the background QFT with a standard Planckian
spectrum whose
temperature is proportional to its acceleration. Thus, even if the
background QFT is originally in its vacuum state the detector can
get excited just because it takes a non--trivial, non--inertial motion
in flat space--time. This is equivalent to the detector being at rest, but
embedded in a heat bath with appropriate temperature.

However, the case of eternal, constant, linear acceleration suffers from
various
subtleties due to the presence of a horizon in the detectors reference
frame. This complicates the study of the Unruh effect due to the necessity
to
input boundary conditions in the detectors proper non--inertial reference
frame
\cite{NFKMB}. This leads one to questionable conclusions
about existence of the Unruh effect \cite{NKMB} and even of Hawking
radiation \cite{Belinski}. We will come back to the discussion
of these concerns below in section 3.
As an aside if the Unruh effect manifested itself
{\it only} for eternal (homogeneous) linear accelerating motion
then the effect would not be physical since it is not possible
to arrange for a particle or reference frame to have such a trajectory.
For this reason we are interested in more realistic and experimentally
obtainable
motions. Of particular interest is the detector moving along a
trajectory which eventually brings it back to the same point in
space. In addition we want the motion to be homogeneous in time or nearly
so since this simplifies the calculations in \eq{grule2} and \eq{grule3}.
Purely homogeneous motions are not experimentally realizable,
but there are cases when a real motion is well approximated by a
homogeneous one.

For these reasons homogeneous, circular motion is the best choice
for our study. In particular, there is no
horizon (only a light surface or light radius) for an observer in circular
motion. Additionally homogeneous circular motion gives
a good approximation for the trajectory of particles in accelerator storage
rings.
However, for circular motion the observed radiation does not have a
thermal spectrum. The detector will still get excited, but the
spectrum of the detected particles is not Planckian.

At the end of the day we would like to study a situation which can in
principle be
realized experimentally. To this end we
take an electrically charged particle with spin as our
detector of the Unruh effect. This brings us to
Sokolov--Ternov effect \cite{SokTer} (see \cite{Jackson},
\cite{LL4} and \cite{Zhukovskibook} for reviews).
The Sokolov--Ternov effect describes the partial depolarization
of electrons in a magnetic field in storage rings due to
synchrotron radiation. The relation
between the Unruh and the Sokolov--Ternov effects has been
previously discussed by Bell and Leinaas \cite{BellLeinaas}, and it
was proposed that one might be able to observe the circular
Unruh effect through this depolarization effect on electrons. Our
conclusions are different
in detail from those in \cite{BellLeinaas}, where the authors observe the
Unruh effect
within Sokolov--Ternov one in different circumstances --- with a
non--constant background
magnetic field. (See also reference \cite{Mane} \cite{matsas} for other
difficulties/subtlies associated with observing the circular Unruh effect).
In particular we find that
because the electron has $g \approx 2$ that the characteristic exponential
factor
of the circular Unruh effect is obscured when the background is that of a
constant
magnetic field. To observe this exponential factor one should
use some particle with $g$ significantly different than $2$. The qualitative
explanation of Sokolov-Ternov effect goes as follows:
\footnote{A more detailed explanation is given in
sections 3 and 4 below. The development in this section is a summary
of \cite{Jackson}.}

It is well known that electrons in circular motion radiate.
In the quasi--classical approximation the total power radiated (due to the
electron's electric charge) is:
\bqa
I_o = \frac{2\, e^2}{3\, R} \,
\gamma^4 \, \omega_0,\label{totalintens}
\eqa
where $e$ is the charge of the electron; $\gamma = 1/\sqrt{1 - v^2}$;
$\omega_0 = 1/R$ is the angular velocity
of the electron in the ultra--relativistic limit ($v \approx 1$);
$R = \mathcal{E}/e\, H$ is the radius of the electron's orbit;
$\mathcal{E} = m\, \gamma$ is the energy of the
electron; $H$ is the background, constant magnetic
field. The quasi-classical approximation is valid
when the electron is ultra--relativistic and we can neglect both
quantization of its motion and back-reaction to the photon
radiation.

Electrons have two energy levels in an
external constant magnetic field: with their spins along and
against the direction of the magnetic field. Hence, electrons can
also radiate via flips of their spins. The ratio of the
power radiated due to the motion of the electron and due to
its spin flip radiations is \cite{Jackson}:
\bqa
\frac{I_{sf}}{I_o} = 3\, \left(\frac{\gamma^2}{m\,
R}\right)^2 \,\left(1 \pm \frac{35\,
\sqrt{3}}{64}\right),\label{ratio}
\eqa
where $I_{sf}$ is the power of the spin flip radiation, and the ``$\pm$''
signs on the RHS correspond to the spin flips with a
decrease/increase of the spin energy, respectively. One can
define a critical $\gamma$--factor as $\gamma_c = \sqrt{m\,R}$.
For characteristic values one finds $\gamma_c \approx 10^7$,
while $\gamma \approx 10^5$. Hence,
the value (\ref{ratio}) is of the order of $10^{-8}$. Thus, only a
very small fraction of the radiation energy is due to the
spin flip. Because of this, the electron beam, which starts
non--polarized, should {\it slowly} become polarized according to the law:
\bqa
\cP(t) = \cP_0 \, \left(1 - e^{-t/\tau_0}\right) =
\frac{8}{5\,\sqrt{3}} \, \left(1 - e^{-t/\tau_0}\right) \approx
0.92 \, \left(1 - e^{-t/\tau_0}\right) . \label{givenpol}
\eqa
The value of equilibrium polarization in the constant external magnetic
field,
$\cP_0$, will be discussed in section 4 (see \eq{polar-4}).
The characteristic time is:
\bqa
\tau_0 = \frac{8}{5\, \sqrt{3}}\, \frac{m^2\, R^3}{e^2\,
\gamma^5}.\label{real}
\eqa
With the characteristic values of $R \sim 1$ km and $\gamma \sim 10^5$ the
time $\tau_0$
is on the order of one hour. At first sight it seems
that the electron beam should become completely polarized
due to the spin flip radiation. However, the spin flip can happen in
both ``directions'' --- either decreasing or increasing the spin
energy. Due to the presence of the latter effect the polarization
is not complete. Of course the total energy of the system --
electron plus radiation -- is not conserved due to the presence of
the external electromagnetic field which drives the electron
along its trajectory.

This sounds very similar to the Unruh effect where the thermal
bath (appearing in the electron's reference system) tends to
depolarize the electrons. However, in this form of the
Sokolov--Ternov effect it is hard to see its relation to the Unruh
effect, because usually the latter contains a characteristic
exponential contribution to the polarization, $\cP_0$, while
\eq{givenpol} does not seem to share this property (in
\eq{givenpol} $\cP_0$ simply equals a constant). The exponential
contribution dominates only when the orbital motion is decoupled
from the internal degrees of freedom, as pointed out in the
discussion two paragraphs after \eq{grule3}. To see the origin of
the exponential part of the effect, one has to consider particles
with arbitrarily gyromagnetic number $g$; \eq{ratio}---\eq{real}
are valid for electrons with $g=2$. To understand what is going
on, let us consider the effect in the inertial reference frame
instantaneously co--moving with the detector. In this frame the
background constant magnetic field is $H' = \gamma \, H$ (we
always consider electrons moving perpendicular to the magnetic
field). The spin degree of freedom can approximately be treated
non--relativistically in this frame. The magnetic moment \bqa
\vec{\mu} = \frac{g}{2} \,\frac{e}{2\,m} \vec{\sigma}, \eqa of
this spin system has two energy levels with an energy difference
\bqa \Delta \mathcal{E} = \left|\frac{g}{2}\right| \, \frac{e\,
H'}{m} = \left|\frac{g}{2}\right| \, \gamma^2 \, \omega_0. \eqa It
seems that in this case {\it the only non-zero} probability rate
is for the spontaneous magnetic dipole transition {\it from the
upper state to the lower one}. The rate is \bqa w_+ =
\frac{4}{3}\, \left(\Delta\mathcal{E}\right)^3 \,
\left|\left\langle 1\left| \vec{\mu} \right| 0\right\rangle
\right|^2 = \frac{2}{3}\, \left|\frac{g}{2}\right|^5 \,
\frac{e^2}{m^2}\, \gamma^6\, \omega_0^3.\label{grule4} \eqa Time
dilation gives a laboratory transition rate reduced by one power
of $\gamma$. With $\omega_0 = 1/R$ and $g = 2$ in the case of
ultra--relativistic electron, \eq{grule4} then leads to the
characteristic time: \bqa \tau_{\infty} = \frac{3}{2}\,
\frac{m^2\,R^3}{e^2\, \gamma^5}.\label{taunaive} \eqa This can be
compared to \eq{real}. Thus, naive arguments give approximately
the correct characteristic time and tell us that eventually all
electrons (i.e. detectors) become polarized, i.e.  $\cP_0 = 1$.

What is wrong with these naive arguments? Why do they not give the
correct answer? This answer is in fact correct for
particles with large $g$, i.e. $g \rightarrow \infty$
or neutral particles. For the latter situation
we should eliminate $g$, $e$ and $R$ in all formulae in favor of the
magnetic field $H$
and magnetic moment $\mu$ by means of $g\, e = 4\, \mu \, m$ and $e\, R =
\gamma\, m/ H$. A clue to this observation can be found from the fact that
a neutral particle with a given magnetic moment can be considered as the
limit
of a particle with infinitesimally small charge and a large (infinite) $g$
factor:
$g = 4\, \mu \, m/e$.

The reason for the relative reliability of the simple arguments for large
$g$ and their
unreliability for small $g$ can be understood by considering the general
features
of the spin motion and the mechanical motion of a charged particle. It is
well
known \cite{LL4} that the magnetic moment of a particle precesses in a
uniform
magnetic field with a frequency (in the laboratory frame):
\bqa
\omega_s = \left[1 + \gamma\, \left(\frac{g-2}{2}\right)\right]\, \omega_0.
\eqa
For a $g$ factor appreciably different from 2, $\omega_s$ becomes very large
compared to
$\omega_0$ for extreme relativistic motion. The relevant quantity is the
number of precessions
during the short time $\Delta t \approx 1/\gamma\,\omega_0$ it takes the
particle to trace
out a segment of path that subtends an angle $\Delta \theta \approx
1/\gamma$ at the
center of the orbit. It is this time interval that is germane to the
emission of
relativistic synchrotron radiation in any given direction (see the
discussion
at the beginning of section 4). For $\gamma \gg 1$ the number
of precessions in $\Delta t$ is $\approx (g-2)/ 4\,\pi$.

For large $g$ the magnetic moment precesses many complete
revolutions during the characteristic time interval $\Delta t$.
This rapidly spinning system has ample time to
establish the two--level system described above and to undergo the
simple magnetic dipole transition, without being influenced
appreciably by the orbital motion. Said another way, the
instantaneously co--moving inertial frame closely approximates the
detector's rest frame for long enough, that the simple
non--relativistic arguments can be applied.

For a $g$ factor of order unity, however, the magnetic moment does not
precess rapidly
enough to ignore the coupling between the spin system and the orbital
motion.
A proper calculation (presented in section 4) shows that \eq{grule4}
receives an exponential correction (for both $w_+$ and $w_-$)
in the large $g$ limit. The correction is proportional to
$e^{- \sqrt{3}(g-2)}/g$.  It is this correction to $w_-$ (making it
non--vanishing) which is usually taken to be the circular Unruh effect.
The exponential factor is equal to 1 if $g=2$ and that is the reason
why the Unruh exponential contribution is usually overlooked
in the standard Sokolov--Ternov calculations.

The organization of the paper is as follows:
Sections 2 and 3 review well known facts about finite temperature
QFT and the reaction of detectors traveling on some given trajectory.
They do not contain any new observations, but give the needed background
for the following calculations and discussion. The reader familiar with all
the standard facts of finite temperature Wightman (Green) functions,
reaction
of detectors to a heat bath and the standard derivation of the Unruh
effect can immediately proceed to section 4, where we discuss
more realistic and more experimentally realizable situations.
Section 5 gives a summary and discussion of the results.

\section{Detector in a heat bath}

In this section we consider the static detector in the heat bath --
background QFT at finite temperature.
The probability for the detector to become excited
or to radiate is given by \eq{grule3}. To perform the integration in this
formula we take, as our example, the two--point correlation
function for the field theory
of one massless real scalar field:
\bqa
G_T(x,\, x') = \frac12 \, {\rm Tr} \, \left[
\frac{e^{-\frac{\hat{H}}{T}}}{Z_{tot}}\,\left\{\phi(x),^{\phantom{\frac12}}
\phi(x')\right\}\right],
\eqa
where  $x$ and $x'$ are four--coordinates: $x = (t, \,
\vec{r})$; $\{,\}$ is the symmetrized product of the fields.
We take the symmetrized product of the fields in the definition of the
two--point function because in this form its connection to
the Unruh effect is more apparent. For the standard, thermal Green functions
given by Keldish \cite{LLX} the answers
are qualitatively similar in the following sense: the equilibrium
polarization for both the Unruh effect and the heat bath are the same but
the way in which the system reaches equilibrium is different \cite{Boyer}.

We Fourier expand the field as follows:
\bqa
\phi (t, \, \vec{r}) = \int \frac{d^3 p}{2\, \pi}\,
\frac{1}{\sqrt{\omega}}\,\left.
\left\{a(\vec{p})^{\phantom{\frac12}}
\exp{\left[-{\rm i}\left(\omega t -
\vec{p} \cdot \vec{r}\right)\right]} +
{\rm c.c.}\right\}\right|_{\omega = |\vec{p}|}
\eqa
and plug this expression into:
\bqa
G_T(x, \, x') = \dots \sum_{n_{p_1} = 0}^{\infty} \sum_{n_{p_2} =
0}^{\infty}
\dots \frac{e^{-\frac{\mathcal{E}_n(p_1)}{T}}}{Z_1}\,
\frac{e^{-\frac{\mathcal{E}_n(p_2)}{T}}}{Z_2} \dots
\nonumber \\
\times \left\langle\dots n^{\phantom{\frac12}}_{p_1},\,
n_{p_2} \dots \left|\left\{\phi(x),^{\phantom{\frac12}}
\phi(x')\right\} \right|\dots n_{p_1}, \, n_{p_2} \dots
\right\rangle,\label{backinto}
\eqa
where $\mathcal{E}_n(p) = \omega(p) \, (n + 1/2)$
and the state $\langle\dots n_{p_1},\,
n_{p_2} \dots|$ contains $n_{p_1}$ quanta of wave vector $\vec{p}_1$,
$n_{p_2}$ quanta of wave vector $\vec{p}_2$, etc. We now obtain:
\bqa
G_T(x, \, x') = \int \frac{d^3 p}{4\, \pi^2} \, \frac{1}{\omega}\,
\sum_{n=0}^{\infty}
\frac{\exp{\left[- \frac{\omega \,\left(n + \frac12\right)}{T}\right]}}{Z}
\left.
\left\{\left(n + \frac12\right)\, \exp{\left[-
{\rm i}\, p
\left(x - x'\right)\right]} + {\rm c.c.}\right\}
\right|_{\omega = |\vec{p}|}.
\eqa
Here the symmetrized product of $a$ and $a^+$ has been used and
\bqa
Z = \sum_{n=0}^{\infty} \exp{\left[- \frac{\omega \,
\left(n + \frac12\right)}{T}\right]}.
\eqa
Using the relation
\bqa
\coth\left(\frac{\omega}{2\, T}\right) = 2\,
\frac{\sum_{n=0}^{\infty} \left(n + \frac12\right)
\exp{\left[- \frac{\omega \,\left(n +
\frac12\right)}{T}\right]}}{\sum_{n=0}^{\infty}
\exp{\left[- \frac{\omega \,\left(n + \frac12\right)}{T}\right]}} =
\frac12 + \frac{1}{e^{\frac{\omega}{T}} - 1},
\eqa
we can represent the two--point function as follows:
\bqa
G_T(x, \, x') = \int \frac{d^3 p}{4\, \pi^2}
\,\coth\left(\frac{\omega}{2\, T}\right)\,\left.
\frac{\cos\left\{p \left(x - x'\right) \right\}}{\omega}
\right|_{\omega = |\vec{p}|}
\nonumber \\
= \int \frac{d^4 p}{(2\, \pi)^4} \,
\coth\left(\frac{\omega}{2\, T}\right)\,
\frac{\exp\left\{-{\rm i}\, p \left(x - x'\right)
\right\}}{p^2 - {\rm i}\, \epsilon}
\nonumber \\
= T \, \sum_{n=-\infty}^{+\infty} e^{- {\rm i}\, 2\, \pi
\, n \, T \, (t-t')} \int \frac{d^3 p}{(2\, \pi)^3} \frac{\exp\left\{-
{\rm i}\, \vec{p} \left(\vec{r} - \vec{r}'\right)
\right\}}{\left(2\, \pi \, n \, T\right)^2 - \vec{p}^2 - {\rm i}\,
\epsilon},\label{int}
\eqa
where $N(\omega) \equiv \coth\left(\omega/2\,T\right)
= 1/2 + 1/(e^{\frac{\omega}{T}} - 1)$ is the
number of states at the energy level $\omega$. The $1/2$ term in the sum
comes from
the zero modes, while the other contribution is the standard Planckian
spectrum.
In \eq{int} the last expression is obtained after integration over $\omega$
in the
complex plane around the poles of the $\coth$ function. To perform the
integration we use the fact that near its poles the $\coth$ function behaves
as:
\bqa
\coth\left(\frac{z}{2\,T}\right) \approx \frac{2\,T}{z - 2\,
\pi \, {\rm i}\, n \,T}, \quad n \in Z.
\eqa
The last expression in \eq{int}
contains the well known sum over the Matsubara frequencies.
However, we have written it in the Minkowski signature, because
we are considering {\it dynamical} fields at finite temperature. The
justification of the use of the Matsubara approach to describe the
dynamics near thermal equilibrium can be found in \cite{LLX}. We have
presented the derivation of the two--point correlation function and written
out all the components in the final expression of \eq{int} to show
the phenomenon which we are studying from various different perspectives.

Evaluating the integrals in \eq{int} gives:
\bqa
G_T(x,\, x') \propto \frac{T}{|\vec{r} - \vec{r}'|} \,\left\{
\coth\left[\pi \,T \left((t - t') + |\vec{r} - \vec{r}'|^{\phantom{\frac12}}
\right)\right]
 - \coth\left[\pi \,T \left((t - t') - |\vec{r} -
\vec{r}'|^{\phantom{\frac12}}\right)\right]
\right\}.
\eqa
It is easy to see from this expression that when $T\to 0$ we obtain
$G_0(x,\, x') \propto 1/(|t-t'|^2 - |\vec{r} - \vec{r}'|^2)$, while
for finite $T$ if we take the limit $\Delta r = |\vec{r} - \vec{r}'| \to 0$
then $G_T(t,\,t') \propto T \, \frac{d\coth[\pi \,T\,(t - t')]}{dt}
\propto T^2/\sinh^2[\pi\, T\,(t-t')]$.

To obtain the probability rate in question we have to substitute
some particular trajectory ${\vec r}(t)$ into this thermal Wightman function
and then
substitute the resulting expression into \eq{grule3}. For a static detector
$x = (t,0,0,0)$ we obtain ($\tau = t - t'$ and $\Delta\mathcal{E} =
\mathcal{E}_u - \mathcal{E}_d$):
\bqa
w_{\mp}(t) \propto \lim_{\epsilon \to 0} q^2 \,
\left|\left\langle 0\left|
\mu^{\phantom{\frac12}}(0)\right|1\right\rangle\right|^2 \,
\int_{-\infty}^{+\infty} d\tau \, e^{\mp
{\rm i}\, \Delta\mathcal{E}\, \tau}\,
\frac{T^2}{\sinh^2\left[\pi \,T \,
\left(\tau - {\rm i}\, \epsilon\right)\right]} ~,\label{thermal}
\eqa
where we have taken the standard regularization of the $\tau = 0$
singularity
of the propagator. The integrand in \eq{thermal} has poles along imaginary
$\tau$ axis corresponding to zeros of the $\sinh$ function. They are
at $\tau = n\, {\rm i}/T + {\rm i} \, \epsilon$, $n \in Z$.
We are in general interested only
in the first, non-trivial pole ($n=\mp1$),
for it is this pole which gives the characteristic Boltzmann
exponential factor in $w_{\mp}$.
However, the integral in \eq{thermal} can be evaluated
exactly (unlike more general cases) by using the following contour (see e.g.
\cite{BellLeinaas}):
Go along the real axis from $\tau = - \infty$ to $\tau = + \infty$;
at $\tau = \pm \infty$ take short vertical lines up to Im$\tau = 2\pi/T$;
take
the the return path along the line Im$\tau = 2\pi/T$. This is a rectangular
contour of infinite length and height $2\pi/T$. Using the result for the
integral
with such a contour we find $w_{\mp}$ and that
the equilibrium ``polarization'' (the ratio of the number of states in the
upper level to those in the lower levels) in the heat bath is:
\bqa
\cP_0 = \frac{w_+ - w_-}{w_+ + w_-} = -
\frac{1 - \exp\left(\frac{\Delta \mathcal{E}}{T}\right)}{1 +
\exp\left(\frac{\Delta\mathcal{E}}{T}\right)} \rightarrow
1 - \exp\left(-\frac{\Delta \mathcal{E}}{T}\right), \quad {\rm if}\quad
\Delta \mathcal{E} \gg T.\label{therpol}
\eqa
We have not discussed the characteristic time for reaching the equilibrium.
This
is contained in the pre--exponential part of $w_{\pm}$ and
properly should be calculated using Keldish
approach \cite{LLX}.

The polarization in \eq{therpol} is not unity since the
background QFT is not in a vacuum state and, hence,
has excitations which can be absorbed by the detector.
Thus, a static detector in a heat bath can
become excited by absorbing energy from the heat bath.

\section{Unruh effect}

In this section we apply \eq{grule3}, with the
background QFT initially in its ground state, (i.e. $T=0$),
but with the detector moving along some
given trajectory. We want to determine if the
detector has a non--zero excitation rate, $w_-$, as it
moves along the trajectory in the
background QFT, which we take to be (for simplicity)
one massless, real scalar field. A more realistic
situation will be studied in section 4.

Under these conditions \eq{grule3} takes the form:
\bqa
w_{\mp}(t) \propto  \lim_{\epsilon\to 0} q^2 \,\left|\left\langle 0\left|
\mu^{\phantom{\frac12}}(0)\right|1\right\rangle\right|^2 \,
\int_{-\infty}^{+\infty} d\tau \times \nonumber \\ \times \frac{e^{\mp {\rm
i}\,
\Delta\mathcal{E}\,
\tau}}{\left|x_0(t+\tau/2) -
x_0(t-\tau/2) - {\rm i}\, \epsilon\right|^2 -
\left|\vec{x}(t+\tau/2) -
\vec{x}(t-\tau/2)\right|^2},\label{explic}
\eqa
where $\tau = t - t'$ is
the detector's proper time and $\Delta\mathcal{E} = \mathcal{E}_u -
\mathcal{E}_d$.
We have explicitly substituted into
\eq{grule3} the Wightman function of a four--dimensional free
massless scalar field at $T=0$. To obtain the rate in question we
have to substitute into \eq{explic} whichever trajectory, $x(t)$,
we are interested in. In general $w_{\mp}$ explicitly depends on time
$t$. However, for homogeneous motions this dependence disappears,
greatly simplifying the calculation of the integral in \eq{explic}.

\subsection{Motion with constant velocity}

In the case of motion with constant velocity $v$ the
trajectory is $x(t) = (\gamma \,t, \gamma \, v\, t, \, 0, \, 0)$, with
$\gamma = 1/\sqrt{1 - v^2}$ and $t$ is the detector's proper time.
Inserting this trajectory into \eq{explic}, we obtain:
\bqa
w_{\mp}(t) = w_{\mp} \propto  \lim_{\epsilon\to 0}\,q^2\,\left|\left\langle
0\left|
\mu^{\phantom{\frac12}}(0)\right|1\right\rangle\right|^2 \,
\int_{-\infty}^{+\infty} d\tau \, \frac{e^{\mp {\rm i}\,
\Delta \mathcal{E}\,
\tau}}{\left(\gamma\,\tau - {\rm i}\, \epsilon\right)^2 - \left(\gamma \,
v\, \tau\right)^2}.\label{constvel}
\eqa
In this case $w_{\mp}$ does not depend on $t$. This happens because the
motion is homogenous. If instead of considering the phenomenon
from the point of view of the co--moving reference frame we had considered
the laboratory
reference frame, $w_{\mp}$ would be changed by a factor of $\gamma$ because
of time dilation
(see the factor of $\sqrt{\dot{x}_\mu^2}$ in \eq{bla}). In \eq{constvel} we
are
using the standard regularization of the Wightman function which
shifts the pole at $\tau = 0$ to the upper complex half--plane.
As we will see in a moment such a regularization is necessary to have
a non--zero probability for the spontaneous radiation of the detector
if it is originally in the excited state.

The integral in \eq{constvel} is taken using contour
integration in the complex $\tau$ plane. Since $\mathcal{E}_u >
\mathcal{E}_d$,
the integral in \eq{constvel}, for $w_-$ uses a contour which is closed
with a large semi-circle in the lower complex half--plane. This contour is
denoted
by $C_-$. For $w_+$ the contour is closed with a large semi-circle in the
upper
complex half--plane, and is denoted by $C_+$. It is this choice of the
contours
which we will use for $w_{\mp}$ everywhere below.

The expression in the denominator of the integrand in \eq{constvel}
has two zeros at
\bqa
\tau_{\pm} = {\rm i} \, \epsilon \, \sqrt{
\frac{1 \pm v}{1 \mp v}}.
\eqa
Hence
\bqa
w_{\mp} \propto \lim_{\epsilon \to 0} \,q^2 \,\left|\left\langle
0\left| \mu^{\phantom{\frac12}}(0)\right|1\right\rangle\right|^2 \,
\oint_{C_{\mp}} d\tau \, \frac{e^{\mp
{\rm i}\, \Delta \mathcal{E}\, \tau}}{\tau_+ - \tau_-} \,
\left(\frac{1}{\tau - \tau_+}
 - \frac{1}{\tau - \tau_-}\right).
\eqa
Evaluating the integrals gives
\bqa
w_- &=& 0, \nonumber \\
w_+ & \propto & q^2 \,\left|\left\langle
0\left| \mu^{\phantom{\frac12}}(0)\right|1\right\rangle\right|^2 \,
\Delta\mathcal{E},
\eqa
because $C_-$ does not enclose any poles, while $C_+$
does. This result is independent of the velocity
and, hence, is valid for the static detector.
The physical meaning of the above result is as follows:
If the detector moves with constant velocity in the vacuum of a QFT
there is zero probability for it to get excited, $w_- = 0$. However, if the
detector
was originally in the excited state, there is a non--zero probability for it
to radiate spontaneously, $w_+ \neq 0$.

\subsection{Motion with constant linear acceleration}

We proceed to the case of constant (in direction
and magnitude) acceleration $a$.
This time $x(t) = \left(\frac{1}{a}\sinh\left[a\, t \right], \,
\frac{1}{a}\cosh\left[a\, t \right], \, 0, \, 0\right)$,
where $t$ is the detector's proper time. Substitution of this expression
into \eq{explic} gives:
\bqa
w_{\mp} \propto \lim_{\epsilon\to 0}\,q^2 \,\left|\left\langle 0\left|
\mu^{\phantom{\frac12}}(0)\right|1\right\rangle\right|^2 \,
\oint_{C_{\mp}} d\tau \, e^{\mp {\rm i}\,
\Delta\mathcal{E}\, \tau} \, \frac{a^2}{
\sinh^2\left[\frac{a}{2}\, \left(\tau - {\rm i} \,
\epsilon\right)\right]}. \label{lineac}
\eqa
Again the rate $w_{\mp}$ does not depend on the time because the
motion is homogenous. In \eq{lineac} the standard
regularization of the Wightman function is used (see \cite{Schlicht} for a
discussion of this point).

The expression in \eq{lineac} is essentially the same as
the heat bath formula \eq{thermal}. Therefore a detector moving with
constant
acceleration in the background QFT (which is originally in the vacuum state)
gets excited. In addition the detector sees particles with the standard
Planckian
distribution. The temperature of the accelerating case can be
read off by comparing \eq{lineac} and \eq{thermal} and is \cite{Unruh}:
\bqa
T = \frac{a}{2\, \pi }.
\eqa
In the previous section it was shown that the static
detector in a heat bath gets excited
due to the absorption of thermal excitations from the background
QFT. This raises the question as to why the accelerating detector
becomes excited if it is in a vacuum state of the QFT? The general physical
explanation is that both the detector and QFT are excited at the
cost of work performed by the force driving the detector along its
trajectory. In greater detail this can be explained as follows:
The Hamiltonian of the background QFT in the detector's co--moving,
non--inertial
reference frame has negative energy eigenstates \cite{Leinaas}. The detector
can radiate
these negative energy eigenstates which then excites the detector. The
energy for
this comes from the external driving force, i.e. the system is not closed.
We can explain the negative energy eigenstates by looking at the Wightman
function which was used in the derivation of \eq{lineac} or \eq{constvel}
under different types of motion:
\bqa
G\left[x(\tau),^{\phantom{\frac12}} x(0)\right] \propto
\frac{1}{\left|x(\tau)^{\phantom{\frac12}} - \,\,\,
x(0)\right|^2} \propto \int \frac{d^3 p}{4\,
\pi^2} \, \frac{1}{\omega}\, \left. \exp\left\{ - {\rm i}\,
p \left[x(\tau) - x(0)\right]
\right\}\right|_{\omega = |\vec{p}|}.
\eqa
First consider constant velocity  motion where
$x(t) = (\gamma \,t, \gamma \, \vec{v}\,t)$. Plugging this representation of
the
Wightman function with this trajectory $x(t)$
into \eq{explic}, we obtain:
\bqa
w_{\mp} \propto  q^2 \,\left|\left\langle
0\left| \mu^{\phantom{\frac12}}(0)\right|1\right\rangle\right|^2 \,
\int_{-\infty}^{+\infty} d\tau \, \int \frac{d^3 p}{4\, \pi^2} \,
\frac{1}{\omega}\, \left.\exp\left\{- {\rm i}\,
\left[\pm \Delta\mathcal{E}^{\phantom{\frac12}} +
\gamma\left(\omega - \vec{p} \cdot \vec{v}\right)\right]\,
\tau \right\}\right|_{\omega = |\vec{p}|}.
\eqa
Taking the integral over $\tau$ first, gives an energy conserving
$\delta$--function as the integrand of the $d^3 p$ integration
\bqa
\delta\left[\pm\Delta\mathcal{E} +
\gamma^{\phantom{\frac12}} \left(\omega -
\vec{p}\ \cdot \vec{v}\right)\right]. \label{deltae}
\eqa
This function vanishes for the upper ($+$) sign
in its argument. For the $+$ sign the argument is always greater than
zero, because $\mathcal{E}_{u} > \mathcal{E}_{d}$ and $|\vec{p}| = \omega >
\vec{p} \cdot \vec{v}$, because $v < 1$ \cite{BirelDavis}.
Hence, in the case of constant velocity motion $w_- = 0$,
but $w_+ \neq 0$, which follows just from energy
conservation.

If the detector moves in an environment
where its velocity is greater than the speed of light, it
can produce Cherenkov type radiation. In this case
it is possible to see that now the
$\delta$--function in \eq{deltae} can be non--zero even for the case
when the sign in front of $\Delta\mathcal{E}$ is the ``+'' sign, i.e. the
argument
in \eq{deltae} can equal zero for some angles between $\vec{p}$ and
$\vec{v}$.
This demonstrates the anomalous Doppler effect \cite{Ginzburg},
which describes the well understood phenomenon that
in such circumstances the detector can both radiate and have its internal
degrees
of freedom excited. A similar thing happens in the case of accelerated
motion.

In the linear acceleration case we find
\bqa
w_{\mp} & \propto & q^2 \,\left|\left\langle 0\left|
\mu^{\phantom{\frac12}}(0)\right|1\right\rangle\right|^2 \,
\int_{-\infty}^{+\infty} d\tau \, \int \frac{d^3 p}{4\, \pi^2} \,
\frac{1}{\omega} \nonumber\\
& \times & \left. \exp\left\{- {\rm i}\, \left(\pm
\Delta\mathcal{E} \tau + \frac{\omega}{a}\, \sinh\left(a
\tau\right) - \frac{\vec{a} \cdot \vec{p}}{a^2} \,
\left[\cosh\left(a \tau\right)- 1\right]\right)
\right\}\right|_{\omega = |\vec{p}|}.\label{39} \eqa After the
integration over $\tau$ one will {\it not} obtain zero regardless
of the sign chosen for $\Delta\mathcal{E}$. Thus, in this case
energy conservation allows the detector to absorb positive energy
states or, equivalently radiate negative energy states. To see
this explicitly one has to take the integral over $\tau$ in the
saddle point approximation (if $\Delta \mathcal{E}/a \gg 1$) and
include contributions of all saddle points of $\tau$, which are
related to the pole contributions in \eq{lineac}. A clearer
physical picture for the appearance of the radiation in the
non--inertial reference frame can be obtained in the
quasi--classical quantization scheme for relativistic particles in
curved stationary backgrounds \cite{AkhSing}. In the latter case
it is straightforward to see that the radiation appears due to
polarization of the vacuum in strong gravitational background
fields (see \cite{Indian} for a more detailed discussion on
this issue).

All well and good, but as mentioned in the
introduction the case of constant homogeneous linear
acceleration is not possible to arrange in reality --
one can not have an eternally accelerating detector.
If one does consider a more realistic motion (e.g. a
stationary detector which accelerates for a finite time
and then moves with constant velocity)
the initial and final conditions increase the difficulty of the
analysis making it much harder (impossible) to
get a clear physical picture of what is going on.
In particular it is not clear whether or not there will be a non--trivial
saddle point contribution as in
\eq{39} if the acceleration is over a finite time.

Moreover, we have considered the phenomenon in question from the
point of view of non--inertial, co--moving reference system. If
instead we study the phenomenon from the point of view of the
laboratory, inertial reference frame then the trajectory is $x(t)
= (t, \, \sqrt{1/a^2 + t^2}, \, 0, \, 0)$. In addition the factor
$\sqrt{\dot{x}_\mu^2}$ in \eq{bla} is non--trivial. Now the motion
does not look homogeneous and $w_{\mp}$ seems to explicitly depend
on time, which again makes the study of the phenomenon difficult
(impossible).

For these reasons we turn our attention to circular
motion. We will consider homogeneous circular motion, i.e. eternal
circular motion with no starting or stopping. However we will show
that homogeneous circular motion is a good approximation for real
circular motion with a starting and stopping time. In the next
section we will discuss under what conditions one could observe
the circular Unruh effect using charged particles with spin as the
detectors.

\subsection{Motion with constant circular acceleration}

If the detector performs homogenous, circular motion \cite{letaw} with
radius $R$ and with the angular velocity $\omega_0$, then
$x(t) = \left(\gamma^{\phantom{\frac12}} t,\, R\, \cos\left[\gamma
\, \omega_0 \, t\right], \, R\, \sin\left[\gamma \, \omega_0 \,
t\right], \,0\right)$, $\gamma = 1/\sqrt{1 - R^2 \, \omega_0^2}$
and $t$ is the detector's proper time.

Inserting this trajectory into \eq{explic}, we obtain:
\bqa
w_{\mp} \propto \lim_{\epsilon\to 0} \,q^2 \,\left|\left\langle
0\left| \mu^{\phantom{\frac12}}(0)\right|1\right\rangle\right|^2 \,
\oint_{C_{\mp}} d\tau \, \frac{e^{\mp {\rm i}\,
\Delta\mathcal{E}\, \tau}}{\left[\gamma\,
\left(\tau - {\rm i}\, \epsilon\right)\right]^2 - 4\, R^2 \,
\sin^2\left[\frac{\gamma \, \omega_0}{2}\, \tau\right]}. \label{cirkmot}
\eqa
Again $w_{\mp}$ does not depend on $t$, because the motion
is homogeneous. Also the integration is over the proper time, but
since circular motion has a simple relationship between proper time, $\tau$,
and laboratory time ($\tau_L = \gamma \, \tau$) one can easily
change to the laboratory frame as in the case of the motion
with constant velocity. This is an important
difference between circular and linear acceleration
which makes the analysis of the circular case more straightforward.
Furthermore, in the case of circular motion one can easily transform
the expression in \eq{cirkmot} from the laboratory reference frame to either
the inertial {\it instanteneously}, co--moving reference frame or the
non--inertial, co--moving
reference frame. Such transformations can not be so easily done in
the case of linearly accelerating motion.

The integrand in \eq{cirkmot} has poles in the complex
$\tau$ plane. In particular it has poles similar in nature
to those in the Wightman function for a heat bath or for
linear acceleration, which lead to a Boltzmann type
exponential contribution to $w_{\mp}$. However, there are also
pre--exponential contributions to $w_{\mp}$ for circular motion
which spoil the exact thermal behavior.

Following \cite{BellLeinaas} let us compare the Wightman functions
for the cases of linear and circular acceleration:
\bqa
G_L(\tau) & \propto & \frac{a^2}{
\sinh^2\left[\frac{a}{2}\, \left(\tau - {\rm i} \, \epsilon\right)\right]}
\approx \frac{1}{\left(\tau - {\rm i} \, \epsilon\right)^2} \, \left(1 +
\frac{1}{12}\,
\left[a\, \tau\right]^2 + \frac{1}{360}\, \left[a\, \tau\right]^4 +
\dots\right)^{-1}, \nonumber \\
G_C(\tau) & \propto & \frac{1}{\left[\gamma\,
\left(\tau - {\rm i}\, \epsilon\right)\right]^2 - 4\, R^2 \,
\sin^2\left[\frac{\gamma \, \omega_0}{2}\, \tau\right]} \nonumber \\
& \approx & \frac{1}{\left(\tau - {\rm i} \, \epsilon\right)^2} \, \left(1 +
\frac{1}{12}\, \left[a\, \tau\right]^2 - \frac{1}{360 \, v^2}\,
\left[a\, \tau\right]^4 + \dots\right)^{-1},
\eqa
where for the case of circular motion the velocity is $v = \omega_0\,
R\,\gamma$ and
the acceleration is $a = \gamma^2\, \omega_0^2 \, R$ in the co--moving
frame. Note that the difference between $G_L(\tau)$ and $G_C(\tau)$
appears only in the third term on the RHS of both equations.

The integral in \eq{cirkmot} can not be done exactly unlike the
linear acceleration case (see \cite{miller} for a semi-analytical
study of this integral in various limiting regimes).
However, as we show in the next section
the circular case simplifies for large $\gamma$. Assuming that the energy
splitting
is not too small (i.e. $\Delta\mathcal{E} > a$) we can approximate the
Wightman function by
\bqa
G_C(\tau) \approx
\frac{1}{(\tau - {\rm i}\, \epsilon)^2} \, \left(1 + \frac{1}{12}\,
\left[a\, \tau\right]^2\right)^{-1}. \label{appgreen}
\eqa
The integral in \eq{cirkmot} can now be computed with the result
\bqa
w_- & \propto & q^2 \,\left|\left\langle
0\left| \mu^{\phantom{\frac12}}(0)\right|1\right\rangle\right|^2
\, a \, e^{- \sqrt{12}\,
\frac{\Delta\mathcal{E}}{a}},
\nonumber \\
w_+ & \propto & q^2\,\left|\left\langle
0\left| \mu^{\phantom{\frac12}}(0)\right|1\right\rangle\right|^2
\, a \, \left(e^{- \sqrt{12}\,
\frac{\Delta\mathcal{E}}{a}} +
4\, \sqrt{3} \,\frac{\Delta\mathcal{E}}{a}
\right).\label{wwpm}
\eqa
The exponential contribution comes from the
non--trivial pole in (\ref{appgreen}) at $\tau = \pm {\rm i}\, 2\sqrt{3}/a$.
The second contribution to $w_+$ comes from the trivial
pole at $\tau = 0$.

For $\Delta\mathcal{E} \gg a$, the equilibrium population of the
upper level relative to the lower level is \bqa \cP_0 = \frac{w_+
 - w_-}{w_+ + w_-} \approx 1 - \frac{1}{4\, \sqrt{3}} \,
\frac{a}{\Delta\mathcal{E}} \, e^{- 2\, \sqrt{3}\,
\frac{\Delta\mathcal{E}}{a}}. \label{eqpopul} \eqa This is exactly
the kind of equilibrium ``polarization'' we will obtain in the
next section when we study the Sokolov-Ternov effect for particles
with large gyromagnetic number $g$. Note that this equilibrium
``polarization'' is not thermal due to the dependence of the
pre--exponential factor on $\frac{a}{\Delta\mathcal{E}} $.
Furthermore, even if we take into account corrections to
\eq{eqpopul} we would not expect to get a thermal spectrum of the
detected particles. Intuition from condensed matter informs us
that the Planckian distribution is strongly related to the form of
the two--point correlation function  in \eq{lineac}. The
two--point function for circular motion given in \eq{cirkmot} has
a drastically different form than that in \eq{lineac}.

The explanation why both $w_+$ and $w_-$ are not zero in this case
is the same as the one given at the end of the previous
subsection. Here we find: \bqa w_{\mp} \propto  q^2
\,\left|\left\langle 0\left|
\mu^{\phantom{\frac12}}(0)\right|1\right\rangle\right|^2
\int_{-\infty}^{+\infty} d\tau \, \int \frac{d^3 p}{4\, \pi^2}
\frac{1}{\omega}\left. \exp\left\{- {\rm i} \left[\left(\pm
\Delta\mathcal{E} + \gamma \,\omega\right) \tau - \vec{p}
\cdot\vec{R}(\tau) \right] \right\}\right|_{\omega =
|\vec{p}|},\label{44n} \eqa where $\vec{R}(\tau) = \left(R\,
\cos\left[\gamma \, \omega_0 \, \tau\right] - 1, \, R\,
\sin\left[\gamma \, \omega_0 \, \tau\right], \,0\right)$. Again
after taking the integral over $\tau$ the resulting expression
under the integral over $d^3p$ is not zero for any choice of sign
in front of $\Delta\mathcal{E}$, which means that the detector can
emit negative energy states, i.e. get excited to $\mathcal{E}_u$.
Again quasi--classics gives a clearer picture of the phenomenon.
According to \cite{Leinaas} (see \cite{AkhSing} for the
qusi--classical study) the detector clicks due to falling
particles to the center (orbiting detector) from the vacuum
fluctuations. See as well \cite{Volovik} for a more detailed
discussion of this issue.

\section{Sokolov--Ternov effect at arbitrary gyromagnetic number and
Unruh effect}

In this section we will use an electrically charged particle with spin
which is undergoing circular motion as our detector for the circular Unruh
effect.
We now give the conditions under which the particle's motion can be
considered classical. We will mostly be discussing electrons, but
we keep the gyromagnetic number $g$ arbitrary. Unlike the previous
cases, in this section we will only consider the laboratory
reference frame and laboratory time.

There are two sources of quantum effects in synchrotron
radiation: (i) the quantization of the electron's trajectory and
(ii) quantum back--reaction under photon emission. The first one is
suppressed if $\omega_0/\mathcal{E} \ll 1$, where $\mathcal{E}$ is the
electron's energy and  $\omega_0 = e\, H/\mathcal{E}$ is the angular
velocity of the electron moving in a background magnetic field
$H$, i.e. $\omega_0$ is the energy splitting between the Landau
levels.

The second source of quantization is defined by the ratio
$\omega_c/\mathcal{E}$, where
\bqa \omega_c = \omega_0 \,
\left(\frac{\mathcal{E}}{m}\right)^3 = \omega_0 \, \gamma^3,
\label{omegac}
\eqa
is the characteristic frequency of the photons emitted in the synchrotron
radiation
\cite{LL4}, \cite{Zhukovskibook}. Here
$m$ is the mass of the electron. If the ratio in question satisfies
\bqa
\frac{\omega_c}{\mathcal{E}} \ll 1
\eqa
then the electron is ultra--relativistic and its motion is
essentially classical.

Another approximation which we adopt comes from the
characteristic features of the radiation in the ultra--relativistic case.
Consider a relativistic electron with
$\gamma = \frac{\mathcal{E}}{m} \gg 1$.
The angular distribution of the radiated power in this
ultra--relativistic limit is proportional to large
powers of
\bqa
\frac{1}{(1 - \vec{n} \cdot \vec{v})}
\eqa
where $\vec{n} = \vec{p}/\omega$ and $\vec{p}$ and $\omega$ are the photon's
momentum and energy \cite{Zhukovskibook}, \cite{LL2}.
Because of the large negative powers of  $1 - \vec{n} \cdot \vec{v}$
the radiation is concentrated in a narrow cone around the direction of
the velocity $\vec{v}$. The angle of the cone is approximately defined by
($v \approx 1$):
\bqa
1 - \vec{n} \cdot \vec{v} = 1 - v \, \cos\theta \approx 1 - v +
\frac{\theta^2}{2}
\approx \frac12 \, \left(\frac{1}{\gamma^2} + \theta^2\right),
\eqa
where $\theta$ is the angle between the velocity $\vec{v}$ and the radiation
direction
$\vec{n}$. Hence, the angle of the radiation cone is
\bqa
\theta < \frac{1}{\gamma} = \frac{m}{\mathcal{E}}.
\eqa
Thus, the radiation in a given direction is formed from the small part
of the trajectory, over which the velocity vector $\vec{v}$ is
rotated by the angle $m/\mathcal{E} \ll 1$. The electron covers this part of
the
trajectory in a laboratory time $\Delta t$ given by
\bqa
\Delta t \, |\dot{\vec{v}}| \approx \Delta t \, \omega_0 \leq
\frac{1}{\gamma} \ll 1.\label{approximour}
\eqa
It is this interval of time which gives the main contribution
to the integrals we calculate below.

\subsection{Synchrotron radiation due to the electric charge}

There are two ways the electron radiates. The first one is the
well known synchrotron radiation of a charged particle. The
interaction Hamiltonian in this case is:
\bqa
\hat{H}_{int} = e \, \vec{A} \cdot \hat{\vec{v}},
\eqa
where the velocity operator is $\hat{\vec{v}} = (-{\rm i}/m)\, \nabla$
and the vector potential $\vec{A}$ is taken to be in the radiation gauge,
$\nabla \cdot
\vec{A} = 0$. The vector potential of a plane electromagnetic
wave with the polarization $\vec{\zeta}$ is
\bqa
\vec{A}(\vec{r}, \, t) = \vec{\zeta} \,
\sqrt{\frac{2\,\pi}{\omega}} e^{-{\rm i}\,\left[
\omega \, t - \vec{p} \cdot \vec{r}\right]} + {\rm c.c.}, \qquad
{\rm where} \quad \omega = |\vec{p}|.\label{vectopot}
\eqa
The radiation probability of a photon with the momentum in
the range $[\vec{p},\, \vec{p} + d\vec{p}]$ is equal to
\bqa
dP = \left|{\rm i} \, \int_{-\infty}^{+\infty}
\left\langle\Psi_f(t)\left| \hat{H}_{int}(t)
\right|\Psi_i(t)\right\rangle \, dt\right|^2 \, \frac{d^3 p}{(2\,
\pi)^3}, \label{genprob}
\eqa
where $\vec{p}$ is the photon
momentum. The integral over the laboratory time $t$ should be
taken over the period of the circular motion \cite{LL4}
\cite{Zhukovskibook}. However, we have taken the integration
region over the whole real line, because the integral is saturated
in the saddle point approximation by the small region of $t$
given in \eq{approximour}.

As in the introduction, we change the integration
variables to $\tau = t - t'$ and $\tau' = t + t'$ and drop the
integral over $\tau'$. In this way we obtain an expression for $dw$ -- the
radiation rate per unit time and per infinitesimal momentum $d\vec{p}$.
The infinitesimal radiation power is $dI_o =
\omega\, dw$. Taking the interaction Hamiltonian which corresponds to
photon emission gives
\bqa
dI_o = \frac{e^2 \, d^3 p}{4\, \pi^2} \,
\int_{-\infty}^{+\infty} d\tau \, \left\langle\Psi_i\left(t +
\frac{\tau}{2}\right)\left| \vec{\zeta} \cdot \hat{\vec{v}}\left(t +
\frac{\tau}{2}\right)\, e^{-{\rm i}\,\left[ \omega \,\left(t +
\frac{\tau}{2}\right) - \vec{p} \cdot \hat{\vec{r}}\left(t +
\frac{\tau}{2}\right)\right]} \right|\Psi_f\left(t +
\frac{\tau}{2}\right)\right\rangle \cdot \nonumber \\
\left\langle\Psi_f\left(t - \frac{\tau}{2}\right)\left|
\vec{\zeta}^* \cdot \hat{\vec{v}}\left(t - \frac{\tau}{2}\right)\,
e^{{\rm i}\,\left[ \, \omega \,\left(t -
\frac{\tau}{2}\right) - \vec{p} \cdot \hat{\vec{r}}\left(t -
\frac{\tau}{2}\right)\right]} \right|\Psi_i\left(t -
\frac{\tau}{2}\right)\right\rangle ,\label{intensity}
\eqa
where the velocity $\hat{\vec{v}}(t)$ and the coordinate
$\hat{\vec{r}}(t)$ are the Heisenberg operators. In the
quasi--classical approximation adopted here these
operators can be substituted by their classical values, i.e.
\bqa
\left\langle\Psi_f(t)\left| \vec{\zeta}^* \cdot
\hat{\vec{v}}(t)\, e^{{\rm i}\,\left[
\omega \, t - \vec{p} \cdot \hat{\vec{r}}(t)\right]}
\right|\Psi_i(t)\right\rangle \rightarrow \vec{\zeta}^* \cdot
\vec{v}(t)\, e^{{\rm i}\,\left[
\omega \, t - \vec{p} \cdot \vec{r}(t)\right]},\label{intver}
\eqa
where $\vec{v}(t) = \dot{\vec{r}}(t)$ and $\vec{r}(t) =
\left(r_1 (t), \, r_2 (t), \, r_3 (t) \right)$ are now classical
velocities and coordinates along the trajectory in laboratory time $t$:
\bqa
r_1(t) = x_0 + \frac{m}{e\, H}\, \sin\left(\frac{e\, H}{\mathcal{E}}\, t +
\varphi\right), \quad r_2(t) = y_0 + \frac{m}{e\, H}\,
\cos\left(\frac{e\, H}{\mathcal{E}}\, t +
\varphi\right).\label{r}
\eqa
These equations for the trajectory are correct
if the electron is ultra--relativistic
($|\vec{v}| \approx 1$) and moving in the
plane perpendicular to the magnetic field $\vec{H}$. The initial
conditions are given by $x_0$, $y_0$, $\varphi$.
Substituting \eq{r} into \eq{intver} and then into \eq{intensity}
and summing over the photon polarizations and integrating over $d \tau$ in
the saddle
point approximation yields, in the large $\omega$ limit \cite{LL4}:
\bqa
\frac{dI_o}{d\omega} \approx \frac{1}{2\, \sqrt{\pi}}\,
\frac{e^3\, \omega_c}{\gamma^2}\,
\left(\frac{\omega}{\omega_c}\right)^{\frac12}
\exp\left\{- \frac{2\, \omega}{3\, \omega_c}\right\}, \quad \omega_c =
\omega_0\,
\gamma^3.\label{energdis}
\eqa
This equation shows that the main contribution to the
radiation comes from the photons with frequency around $\omega_c$.
This confirms the discussion around \eq{omegac}.
If in \eq{intensity} we had instead taken the integrals over both
$\vec{p}$ and $\tau$ in the appropriate approximation we would have obtained
\eq{totalintens}.

\subsection{Synchrotron radiation due to spin flip}

In addition to the radiation associated with its charge the electron
can also radiate via a spin flip transition. The energy distribution of this
radiation is similar to
the one given in \eq{energdis} \cite{Jackson}.
We are interested in the probability rate
of the radiation, which can be obtained from
the relativistic motion of a spin $\vec{s}$ as given by \cite{LL4}:
\bqa
\frac{d \vec{s}}{d t} & = & {\rm i} \, \left[\hat{H}_{int}, \,
\vec{s}\right], \nonumber \\
\hat{H}_{int} & = & - \frac{e}{m}
\, \vec{s}\, \left[\left(\alpha + \frac{1}{\gamma}\right)\, \vec{H} -
\frac{\alpha\, \gamma}{\gamma + 1}\, \vec{v}\,\left(\vec{v} \cdot
\vec{H}\right) - \left(\alpha + \frac{1}{\gamma + 1}\right)\,
\vec{v} \times \vec{E}\right],\label{hinter}
\eqa
where $\alpha = (g-2)/2$ is the magnetic
moment anomaly, $\vec{v}$ is the particles velocity and
$\vec{E}$ is the electric field. Substituting
the vector potential for the outgoing photon from \eq{vectopot}
into \eq{hinter} and then substituting this Hamiltonian into \eq{genprob},
yields the total probability rate (in the quasi--classical approximation of
the
electron's motion)
\bqa
w_{\mp}(t) = \frac{e^2}{4\, \pi^2 \, m^2} \, \sum_{\rm phot.\,
pol.} \int \frac{d^3 p}{\omega} \int_{-\infty}^{+\infty}
d\tau\, \left\langle i\left|s_k^*\left(t +
\frac{\tau}{2}\right)\right|f\right\rangle \left\langle
f\left|s_j\left(t - \frac{\tau}{2}\right)\right|i\right\rangle \cdot
\nonumber \\
W_{km}\left(t - \frac{\tau}{2}\right)\,
\zeta_m^* \, W_{lj}\left(t + \frac{\tau}{2}\right) \, \zeta_l
\exp\left\{-{\rm i}\, \left( \omega \, \tau - \vec{p} \cdot
\left[\vec{r}\left(t - \frac{\tau}{2}\right) - \vec{r}\left(t +
\frac{\tau}{2}\right)\right]\right)\right\}\label{cyclotprob}
\eqa
where we have taken the sum over the photon polarizations and the $W_{il}$
are given by
\bqa
W_{il}(\vec{p},\, \omega, \,t) = \left[\left(\alpha +
\frac{1}{\gamma}\right)\, \epsilon_{ijl}\, p_j - \frac{\alpha\,
\gamma}{\gamma + 1}\, v_i(t) \epsilon_{jml} \, v_j(t)\,
p_m - \left(\alpha + \frac{1}{\gamma + 1}\right)\, \epsilon_{ijl}
\, v_j(t) \, \omega\right] \label{diffop}
\eqa
and $\vec{v}(t) = \dot{\vec{r}}(t)$.
The origin of the subscripts $\mp$ in the LHS of
\eq{cyclotprob} will be explained in a moment.

In \eq{cyclotprob} the integral over $\tau$ should be over the period
of the circular motion. However, we can extrapolate the
integration region to the whole real line, because the integral is saturated
by the small region of values of $\tau$ defined in
\eq{approximour}. It is this approximation which allows us to model
the realistic motion of charged particles in storage rings by homogeneous
circular motion.

To evaluate $w_{\mp}(t)$ further we need to evaluate the
expectation values of the spin operators. From \eq{hinter} with
a constant magnetic field, $H$, the spin operators evolve in
laboratory time according to
\bqa
s_{\pm}(t) & = & s_1(t) \pm {\rm i}\, s_2(t) = s_{\pm}(0)\, e^{\pm {\rm i}\,
\omega_s \, t}, \quad s_3(t) = s_3(0), \nonumber \\
\omega_s & = & \left[1 + \gamma\,
\left(\frac{g-2}{2}\right)\right]\, \omega_0,
\eqa
where $\omega_s$ is the precession frequency of the spin in the external,
constant magnetic field, i.e. it is the energy difference $\mathcal{E}_u -
\mathcal{E}_d$ between the upper and lower spin energy levels.
Because we are interested in spin flip transitions we
take the initial value of the spin ($|i \rangle$) either
along or against the magnetic field and then flip it
($\langle f | \ne \langle i |$). This yields,
\bqa
\left\langle f \left|s_{+}(0)\right|i\right\rangle & = &
\left\langle f \left|\frac14 \, \left(\sigma_1 + {\rm i}\,
\sigma_2\right)\right|i\right\rangle =
 - \frac14 \, \left(1 \mp 1\right), \nonumber \\
\left\langle f \left|s_{-}(0)\right|i\right\rangle & = & \left\langle
f \left|\frac14 \, \left(\sigma_1 - {\rm i}\,
\sigma_2\right)\right|i\right\rangle = \frac14 \, \left(1 \pm
1\right), \nonumber \\
\left\langle f \left|s_z(0)\right|i\right\rangle & = & \left\langle f
\left|\frac12
\, \sigma_3\right|i\right\rangle = 0.
\eqa
In the first two equations the upper sign corresponds to the spin flip which
increases the spin energy, while lower sign decreases the spin energy.
These signs correspond to the signs in \eq{cyclotprob}.
In addition using
\bqa
\sum_{\rm phot.\, pol.} \zeta_m^* \, \zeta_l = \delta_{ml}
 - \frac{p_m\, p_l}{p^2} ,
\eqa
and
\bqa G(\vec{r}, \, t) =  \frac{4\,\pi}{\left(t - {\rm i}\,
\epsilon\right)^2 - \vec{r}^2} = \int \frac{d^3 \vec{p}}{\omega} \left.
\exp\left\{- {\rm i}^{\phantom{\frac12}} \left(\omega \, t - \vec{p} \cdot
\vec{r}\right)\right\}\right|_{\omega =
\left|\vec{p}\right|} ,
\eqa
the probability rates are
\bqa
w_{\mp}(t) & = &\frac{e^2}{4\,
\pi^2 \, m^2} \, \lim_{\epsilon\to 0}
\,\left\langle i\left|s_k^*(0)^{\phantom{\frac12}} \right|f\right\rangle
\left\langle f\left|s_j(0)^{\phantom{\frac12}}\right|i\right\rangle
\oint_{C_{\mp}} d\tau\, \nonumber \\
& \times & \left[\hat{W}_{km}\left(t - \frac{\tau}{2}\right)\, \,
\hat{W}_{mj}\left(t + \frac{\tau}{2}\right) + \left(\alpha +
\frac{1}{\gamma + 1}\right)^2 \, \epsilon_{knm} \, v_n \,
\epsilon_{jli}\, v_l\,\frac{\partial}{\partial r_m} \,
\frac{\partial}{\partial r_i}\right]
\nonumber \\
& \times & \left.\frac{4\, \pi\, e^{\mp {\rm i} \, \omega_s \, \tau}}{
\left(\tau - {\rm i}\, \epsilon\right)^2 - \left[\vec{r} -
\vec{r}'\right]^2}\right|_{r = r\left(t - \frac{\tau}{2}\right),
\,\, r' = r\left(t + \frac{\tau}{2}\right)},
\label{integralevent}
\eqa
where $\hat{W}$ is the differential
operator obtained by substituting  $\vec{p} = {\rm i} \,
\partial/ \partial \vec{r}$ and $\omega = {\rm i} \, \partial/ \partial
t$ into \eq{diffop}. Taking a homogeneous circular trajectory for
$r(t)$ we find that, up to the pre--exponential differential
operator, \eq{integralevent} coincides with \eq{cirkmot}.
$\mathcal{E}_u - \mathcal{E}_d$ is replaced by $\omega_s$, since
now the energy difference comes from the electron's spin in a
constant, background magnetic field. This pre--exponential is the
source of the difference between the standard Sokolov--Ternov
effect (where the detector is coupled to the electromagnetic field)
and the circular Unruh effect (where the detector is coupled to a
scalar field). Usually one takes the Unruh effect as being given
only by the exponential contribution to $w_{\mp}$ as in \eq{wwpm}.

To take the integral in \eq{integralevent} we change the integration
variable to $z = \tau \, \omega_0 \, \gamma$, take the contours, $C_{\mp}$,
defined
in the previous section, use the approximate expression for the Wightman
function
\eq{appgreen} and use the standard integrals
\bqa
\label{standint}
\lim_{\epsilon\to 0} \oint_{C_-} \frac{e^{-{\rm i}\, A \, z}\,dz}{
\left(z - {\rm i}\, \epsilon\right)^n \, \left(1 + \frac{z^2}{12}\right)^m}
&=&
\frac{{\rm i}^n \, e^{- A\, \sqrt{12}\,} \pi
\left(\sqrt{12}\right)^{1-n}}{(m-1)!}
\\
&\times &
\left(\frac{n+1}{2}\right)\left(\frac{n+1}{2} + 1\right)
\dots \left(\frac{n+1}{2} + m - 2\right), \quad m \ge 1,
\nonumber
\eqa
and similarly for $C_{+}$. The non--trivial pole contribution in
\eq{integralevent}
from the integral over
$\tau$ is the same as the saddle point contribution which plays its role if
in
\eq{cyclotprob} one takes the integral over $\tau$ first and then takes
the integral $d^3 p$.
An important point to note is that after substituting the contribution
of the non--trivial pole ($z = \pm {\rm i}\, 2\, \sqrt{3}$) into the
exponent of
\eq{integralevent} we obtain
an expression proportional to $e^{-(1/\gamma + \sqrt{12} \, \alpha)}$.
If $g \approx 2$ (the case of electrons) and $\gamma \gg 1$ the exponential
factor $\approx 1$. This is the reason why its contribution is
usually overlooked in the standard Sokolov--Ternov calculation.

Combining \eq{integralevent}, \eq{appgreen}, \eq{standint} and considering
only
$\alpha > 0$ yields \cite{Jackson}
\bqa
w_{\mp} \approx \frac{1}{2\, \tau_0} \, \left\{F_1(\alpha)\,
e^{-\sqrt{12} \,\alpha} + F_2(\alpha) \mp F_2(\alpha)\right\},\label{FF}
\eqa
where $\tau_0$ is given in \eq{real} and
\bqa
F_1(\alpha) &=& \left(1 + \frac{41}{45}\, \alpha -
\frac{23}{18}\, \alpha^2 - \frac{8}{15}\, \alpha^3 + \frac{14}{15}
\,\alpha^4\right) - \frac{8}{5\,\sqrt{3}}\, \left(1 +
\frac{11}{12}\, \alpha - \frac{17}{12}\, \alpha^2 - \frac{13}{24}
\,\alpha^3 + \alpha^4\right), \nonumber \\
F_2(\alpha) &=& \frac{8}{5\,\sqrt{3}}\, \left(1 +
\frac{14}{3}\, \alpha + 8 \, \alpha^2 + \frac{23}{3}
\,\alpha^3 + \frac{10}{3}\,\alpha^4 + \frac{2}{3}\, \alpha^5\right).
\eqa
One can see the exponential factor in \eq{FF}, which
appears for the same reasons as the one in \eq{wwpm}. This is the Unruh type
contribution. Let us look at the phenomenon in greater detail.
If $g=2$ (i.e. $\alpha = 0$) we obtain the characteristic time
equal to $\tau_0$ and the equilibrium polarization is
given by \eq{givenpol} where:
\bqa
\label{polar-4}
w_{\mp} \approx \frac{5\, \sqrt{3}}{8}\, \frac{e^2\,\gamma^5}{m^2\, R^3}\,
\left(1 \mp \frac{8\, \sqrt{3}}{15}\right) \quad
{\rm and} \quad  \cP_0 = \frac{w_+ - w_-}{w_+ + w_-} = \frac{8}{5 \sqrt{3}}.
\eqa
If instead one considers the $g \gg 1$ limit then one obtains
the simple rate given in \eq{grule4}, as discussed in
the introduction. The first correction to \eq{grule4} in this limit
is due to the $\alpha ^4 / \alpha^5$ term
in $F_1 (\alpha) / F_2 (\alpha)$ as can be seen from the probability rates
in \eq{FF}:
\bqa
w_{\mp} \approx \frac{g^5}{2^6\, \tau_0} \, \left\{\left[\frac{14}{15}
 - \frac{8}{5\, \sqrt{3}}\right] \, \frac{e^{- \sqrt{3}\, (g - 2)}}{g}
+ \frac{16}{15\, \sqrt{3}} \mp \frac{16}{15\, \sqrt{3}} \right\} +
O\left(\frac{1}{g^2}\right).
\eqa
In this limit the equilibrium polarization is
\bqa
\cP_0 \approx \frac{1}{1 + \frac{F_1(\alpha)}{F_2(\alpha)}\,
e^{- \sqrt{3}\, (g - 2)}} \approx 1 - \left[\frac{\rm const}{g}
+ O\left(\frac{1}{g^2}\right)\ \right]\, e^{- \sqrt{3}\, g}.
\eqa
Comparing this result with \eq{eqpopul} we find perfect agreement if we take
into
account that here $\Delta\mathcal{E}/a \approx g/2$ and
we are using the limit $g,\gamma \gg 1$.

\section{Conclusions}

The arguments presented
above show that physically the Sokolov--Ternov and Unruh effects
are essentially the same, i.e. if we study the Sokolov--Ternov
effect for electrons from the point of view of the non--inertial,
co--moving, reference frame then it appears to arise from an Unruh
type radiation. In fact, in the non--inertial co--moving reference
frame the electrons are at rest. In this case the spin flip
transition with the decrease of the spin energy can happen at
least due to the spontaneous radiation. But what is the reason for
the spin flip transition with the increase of the spin energy if
the electrons are at rest? In this note we show that the latter
transition happens due to existence of the Unruh type radiation
existing in the non--inertial co--moving reference frame.

Usually one considers the Unruh effect under the condition
where the motion of the detector is decoupled from its internal degrees
of freedom (as explained in the introduction). In this situation
the Unruh effect manifests itself through the characteristic exponential
factor in the probability excitation rate.
However, in the experimentally realizable situation one has to consider
charged particles with the spin as the detectors for Unruh type radiation.
In this case usually there is no decoupling of the motion of the detector
from its internal degrees of freedom. Only if $g\to\infty$ do we obtain the
decoupling
and then the characteristic exponential factor dominates\footnote{Note that
there
is always the exponential contribution to the Sokolov--Ternov effect, i.e.
even if $g=2$. In the latter case it is proportional to $e^{-1/\gamma}$ and
basically
equal to $1$ in the extreme relativistic limt $\gamma\to\infty$.}.
For general values of $g$ the exponential factor is mixed with
the pre--exponential contributions. In any case,
we have shown that if one accepts the Sokolov-Ternov effect then this
immediately
implies the existence of the circular Unruh effect (and by extension
the linear Unruh effect and also Hawking radiation from
black holes) since both the Sokolov-Ternov and Unruh exponential factor
appear due to the contribution of the non--trivial pole at
$z = - {\rm i}\, 2\,\sqrt{3}$ in \eq{integralevent} and are physically due
to the
fact that detectors (electrons) see exotic radiation in their non--inertial
reference frame.
It just happens that as we change $g$ the way in which the detector couples
to the background
QFT changes as well.

However, the experimental observation of the characteristic exponential
factor is difficult for electrons which have $g \approx 2$
making this factor $\approx 1$. One can consider using a stable enough
particle with $g$
substantially different from 2. The proton with $g \approx 5.6$ seems the
best candidate.
However, since the proton's mass is approximately
2000 times that of the electron the relaxation time $\tau_0$, \eq{real},
will be much larger than the relaxation time for the electron. Changing only
the
mass term in \eq{real} gives a relaxation time on the order of $10^6$ times
longer than for the electron. Also it is much harder to ``cook" protons so
as to
obtain a large enough $\gamma$--factor. Recent values of the
$\gamma$--factor
for protons produced at LHC are only $7 \times 10^3$ . These values of
$\gamma$, $m$ and
$R \sim 4$ km (as at the LHC) give enormous relaxation times making the
the exponential, Unruh type contribution to the polarization impossible to
detect
using protons at present accelerators.

The main reason for interest in the Unruh effect is that it
provides a simple test case for understanding how to properly
quantize a field theory in a background different from Minkowski.
This by extension would allow one to get a deeper understanding of
closely related effects such as black hole radiation. However, in
this paper we have considered the Unruh effect from the point of
view of the spacetime with the Minkowski metric. But it is its
consideration from the point of view of the metric of the
non--inertial, co--moving reference systems (in which the detector
is stationary) which leads one to addressing the problem of
quantizing a field theory in a background different from Minkowski
spacetime. To accomplish this one should change from Minkowski
coordinates, which are seen by the laboratory observer, to Rindler
coordinates, which are seen by the co--moving observer. The
coordinate change is \bqa x_0 = \frac{\rho}{a}\, \sinh{(a\,t)},
\qquad x = \frac{\rho}{a} \, \cosh{(a\,t)}.\label{Rindchange} \eqa
This transforms the Minkowski metric to the Rindler metric \bqa
ds^2 = - (a\,\rho)^2\, dt^2 + d\rho^2 + dy^2 + dz^2.\label{Rind}
\eqa To find the Wightman function for the scalar field in Rindler
space we have to take the Wightman function in Minkowski space and
make the coordinate change \eq{Rindchange}. For the static
trajectory in Rindler space the result is given in \eq{lineac}. We
can study the orbiting observer case in a similar way, and give an
answer as to whether or not a detector in circular motion will
click \cite{louko} (i.e. observe particles/radiation ). From our
analysis in sections 3 and 4 we can answer in the affirmative: the
detector will indeed click! Note that the stationary, classical
motion of the detector is equivalent to the metric being
stationary in the detectors reference frame, i.e. it is in this
case that we can neglect the gravitational back-reaction connected
with the radiation. It is in this approximation that our
conclusions are valid. In particular, only in the case when the
back-reaction can be neglected does the detector moving with
constant, linear acceleration see particles with a Planckian
spectrum. If the motion is not homogeneous, the spectrum is
drastically changed. As we have seen throughout this note, the
spectrum is different from a thermal spectrum even in the case of
different types of homogeneous motion. These observations should
be important toward resolving the black hole information paradox.

We can also use arguments in this paper to understand what is
observed by a detector in the background of a Schwarzschild black
hole with the mass $M$: \bqa ds^2 = - \left(1 -
\frac{2M}{r}\right) \, dt^2 + \frac{dr^2}{\left(1 - \frac{2M}{r}
\right)} + r^2 d\Omega^2.\label{Schwarz} \eqa We want to study the
behavior of a detector at rest with respect to the black hole in
the vicinity of the horizon where the gravitational field is
almost homogeneous (the gravitational field is almost homogeneous
in the small vicinity of any surface with constant $r$). Based on
the equivalence principle we expect that the behavior of such a
detector will be similar to the behavior of the linearly
accelerating detector. To see this connection we make the
following coordinate change in \eq{Schwarz}: \bqa \rho = \sqrt{1 -
\frac{2M}{r}}. \eqa This converts the Schwarzschild metric to \bqa
ds^2 = - \rho^2 \,dt^2 +
\left(\frac{4M}{(1-\rho^2)^2}\right)^2\,d\rho^2 +
\left(\frac{2M}{1-\rho^2}\right)^2\,d\Omega^2. \eqa The horizon is
at $\rho=0$. Hence, in the vicinity of the horizon the metric is
\bqa ds^2 \approx - \rho^2 \,dt^2 + \left(4M\right)^2\,d\rho^2 +
\dots \eqa One can show via a simple coordinate rescaling that
this metric coinsides with the Rindler metric of \eq{Rind} in the
$\rho-t$ space. Thus, a detector which is at rest in the vicinity
of a Schwarzschild black hole will see thermal radiation for the
same reason as a detector at rest in the Rindler metric.

All these arguments give a clearer picture of the conditions under
which an observer will or will not detect particles/radiation and
the character of the radiation. However, there are confusing
\cite{Belinski} aspects of this topic which require a more
detailed investigation of the quantization of fields in
curvilinear coordinates. For example, one would like to know the
details of the complete basis of harmonics in spaces with horizons
and how these basis harmonics transform under various coordinate
changes. The main question one would like to be able to answer in
all these different cases is: ``What is the criteria for the
existence of radiation due to some particular gravitational
background and/or detector motion?"

The presence of a horizon is not a necessary criteria; there is no
horizon for the orbiting observer. In fact, there is a crucial
difference between the Rindler and the orbiting observers. The
Rindler observer has a horizon, since depending on the
acceleration there is some point behind the observer beyond which
even a massless particle can not reach the observer. On the other
hand the orbiting observer can always be reached by particles from
outside his light surface. Note, however, that an observer
accelerating linearly for a finite time does not have a horizon.

Furthermore, it is {\it not} correct to say that any non--inertial
moving observer sees radiation, while any inertial moving observer
does not. Indeed, a free falling detector in the black hole
background does see radiation (e.g. an observer orbiting around a
black hole). It is only the free falling observer in a homogeneous
gravitational field (i.e. with zero Riemann tensor) which does not
encounter radiation. At the same time an observer fixed above a
gravitating body without a horizon (such as the Earth) does not
see any radiation. Of course in the latter case the radiation
would be so small that it could not be excluded experimentally.
However, theoretically there is no means for such an object as the
Earth to create particles and lose mass.

Thus, at this stage we propose the following criteria for the
existence of radiation from a gravitational background and/or
detector motion: a gravitational background and/or detector motion
will have radiation associated with it if there exist negative
energy states for the Hamiltonian of the QFT in the non--inertial
reference frame \cite{Leinaas}. Rephrasing, the criteria is based
on the existence of a non--trivial saddle point contribution in
the analogs of \eq{39} and \eq{44n} for general motions and/or
backgrounds.

\begin{center}
{\bf Acknowledgment}
\end{center}
AET would like to thank S.Dubovski, P.Tinyakov, V.Rubakov, I.Polubin,
N.Narozhny, A.Fedotov, V.Mur and especially
A.Mironov and S.Mane for the valuable discussions.
As well we would like to thank A.Morozov for collaboration
at the initial stage of the work on this project. This work
supported by the following grants: RFBR 04-02-16880 and the Grant from the
President
of Russian Federation for support of scientific schools, and
a CSU Fresno International Activities Grant.

\end{document}